\documentclass[prb,a4paper,twocolumn,showpacs,superscriptaddress]{revtex4}

\usepackage{amsmath}
\usepackage{amssymb}
\usepackage{graphicx}

\begin{document}

\title{Power law dependence of the angular momentum
transition fields in few-electron quantum dots}

\author{E. Anisimovas}
\affiliation{Departement Natuurkunde, Universiteit Antwerpen (Campus
Drie Eiken), Universiteitsplein 1, B-2610 Antwerpen, Belgium}
\author{A. Matulis}
\affiliation{Semiconductor Physics Institute, Go\v{s}tauto 11,
LT-2600 Vilnius, Lithuania}
\affiliation{Departement Natuurkunde, Universiteit Antwerpen (Campus
Drie Eiken), Universiteitsplein 1, B-2610 Antwerpen, Belgium}
\author{M. B. Tavernier}
\affiliation{Departement Natuurkunde, Universiteit Antwerpen (Campus
Drie Eiken), Universiteitsplein 1, B-2610 Antwerpen, Belgium}
\author{F. M. Peeters}
\affiliation{Departement Natuurkunde, Universiteit Antwerpen (Campus
Drie Eiken), Universiteitsplein 1, B-2610 Antwerpen, Belgium}

\date{15 February 2004}

\begin{abstract}
We show that the critical magnetic fields at which a few-electron
quantum dot undergoes transitions between successive values of its
angular momentum ($M$), for large $M$ values follow a very simple
power-law dependence on the effective inter-electron interaction
strength. We obtain this power law analytically from a quasi-classical
treatment and demonstrate its nearly-universal validity by comparison
with the results of exact diagonalization.
\end{abstract}


\maketitle

\section{Introduction}

The field of quantum dots has been evolving under the busy attention of
theorists and experimentalists alike for already almost 20 years by
now.\cite{jacak98,maks00,kouw01,reimann02}
This amount of interest and the enthusiasm is due to the unique blend of
technological advances and fundamental interest. A special role is
assigned to the magnetic field as a versatile tool with which one can
tune the electronic properties of the quantum dots. In
particular, the application of a perpendicular magnetic field aids the
Wigner crystallization,\cite{mat_ssc,reimann00} the formation of strongly
correlated many-body states\cite{jain95,kamilla95}
and induces ground state multiplicity transitions.\cite{merkt91}

From the theoretical side, the `exact' numerical diagonalization
method\cite{mikh3,mikh3b,mikh4,maarten03,yang93,eto97,yang02,reimann00}
is of profound importance as a reference for approximate treatments.
With the presently available computing power this approach has been
successfully applied to compute the electronic structure of few-electron
quantum dots. However, this method also has
sharp limitations since the numerical effort grows exponentially
with the number of electrons. Therefore, the construction, refinement
and application of less demanding approximate approaches based on some
enlightening or visualizing idea is of critical importance.

Among such ideas is the successful introduction of quasiparticles that
came to be known as {\it composite fermions}.\cite{cf,jain90}
This transformation reduces
the initial problem to a much more tractable problem of non-interacting
or, at least, weakly interacting particles. Another fruitful idea is to
gain insight from the classical picture. The electrons in a quantum dot
can be viewed as classical particles vibrating around their equilibrium
positions\cite{bed94,matulis94,me98} not unlike atoms in a molecule or
a solid, while the molecule can rotate as a whole.\cite{maks96}
These developments culminated in the formulation of the rotating electron
molecule approach\cite{yann02,yann03} based on a classical visualization
and successfully competing with the composite fermion model.

In the present work, we call attention to yet another manifestation
of the classical nature of quantum dots in strong magnetic fields.
Namely, we concentrate on the critical magnetic fields, i.~e.\ the
field strengths at which the ground-state angular momentum of the dot
switches between its two successive values. Here, we found that
these critical field values
show a strikingly simple dependence on the effective Coulomb coupling
strength. We compare estimates extracted from an unsophisticated
quasi-classical model with the exact-diagonalization results thereby
demonstrating the viability of such essentially classical paradigms
and presenting one more argument in their favor.

The paper is organized as follows. In Sec.\ II we present the
exact-diagonalization results indicating that the critical magnetic
fields obey a simple power law. In Sections III and IV  we introduce
and solve a simple quasi-classical model, and compare its predictions
to the exact results in Sec. V. The paper ends with a concluding
Sec. VI.

\section{Exact diagonalization}

\begin{figure}[ht]
\includegraphics[width=80mm]{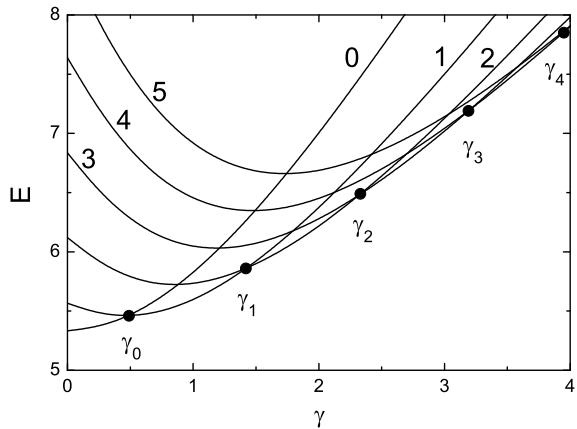}
\caption{\label{fig1}The spectrum of a strongly interacting ($\lambda = 5$)
two-electron parabolic dot as a function of perpendicular magnetic field.
The lowest term of each angular momentum $M = 0 \ldots 5$ is shown. The
Zeeman energy is not included. At the critical fields $\gamma_M$ the
angular momentum switches from $M$ to $M+1$.}
\end{figure}

In Fig.~\ref{fig1} the well-known energy spectrum of two electrons in
a parabolic dot\cite{merkt91} (the Zeeman energy is not included) is
plotted as a function of the perpendicular magnetic field strength $B$.
We work with dimensionless variables so that
the system energy is measured in units of $\hbar\omega_0$ with $\omega_0$
being the characteristic frequency of the confinement potential, while
the magnetic field strength is expressed as the ratio
$\gamma = \omega_c/\omega_0$ of the cyclotron frequency
$\omega_c=eB/mc$ to the above-introduced confinement frequency. The
relative importance of the electron-electron Coulomb interaction is
characterized
by the dimensionless coupling constant $\lambda = a_0/a_B$. Here
$a_0=\sqrt{\hbar/m\omega_0}$ is the characteristic dimension of the
parabolic quantum dot, and $a_B=\epsilon\hbar/me^2$ is the Bohr radius.
Fig.~\ref{fig1} corresponds to a relatively large (i.~e.\ strongly
interacting) dot of $\lambda=5$, and we include the magnetic fields up
to $\gamma = 4$ and angular momenta up to $|M| = 5$. We observe that for
positive magnetic fields the preferred values of the angular momentum are
negative, however, for the sake of convenience we will henceforth use
the symbol $M$ for its absolute values and dispense with the ``$-$'' sign.

The most conspicuous feature of this spectrum is the presence of certain
critical values $\gamma_M$ at which the ground state of the quantum dot
changes the absolute value of its total angular momentum from $M$ to $M+1$.
This feature is brought about by the electron-electron interaction and
persists for any number of electrons in the dot.

\begin{figure}
\includegraphics[width=80mm]{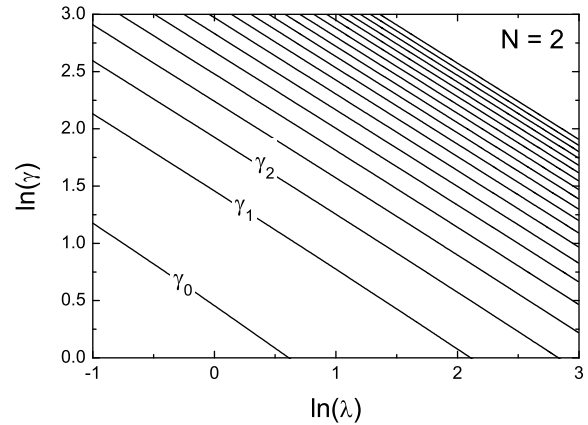}
\caption{Critical magnetic field strength values for two electrons.
The full lines delimit regions characterized by different ground-state
angular momenta starting with $M = 0$ at the lower left corner and
increasing in steps of $\Delta M = 1$ at each consecutive boundary.
Note the monotonic decrease of inter-line distances with increasing
$\lambda$ and $\gamma$.}
\label{fig2}
\end{figure}

\begin{figure}
\includegraphics[width=80mm]{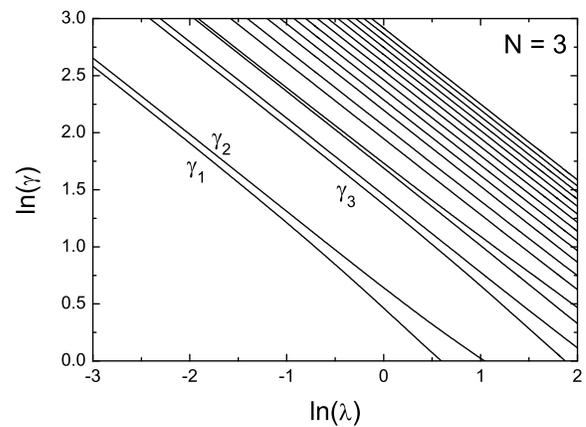}
\caption{The same as in Fig.\ \ref{fig2} but now for three electrons.
An additional phase boundary separating regions of angular momentum
$M = 0$ and $M = 1$ that shows a different behavior is not shown.
Note that widths of some regions corresponding to more stable
states are considerably larger.}
\label{fig3}
\end{figure}

\begin{figure}
\includegraphics[width=80mm]{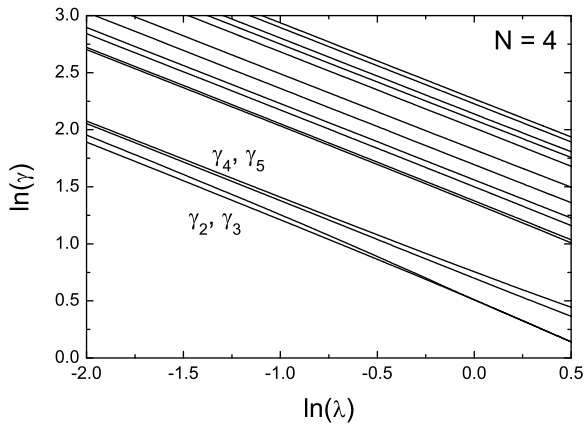}
\caption{The same as in Fig.\ \ref{fig2} but now for four electrons.
 An additional phase boundary separating regions of angular
momentum $M = 0$ and $M = 2$ that shows a different behavior is not
included. The regions corresponding to more stable states are wider.}
\label{fig4}
\end{figure}

We performed exact numerical diagonalization for quantum dots containing
$2$, $3$ and $4$ electrons (see Ref.\ \onlinecite{maarten03} for the
description of the procedure) and calculated the critical magnetic field
values
for various effective Coulomb interaction strengths. For the two-electron
quantum dot we included $\lambda$ values up to $20$, and for the case of
three electrons in the dot the upper limit was set at $\lambda = 10$. The
case of four electrons is numerically more demanding and we restricted
the calculation to values up to $\lambda = 2$. We note that the effective
electron-electron interaction strength can be easily tuned by varying the
strength of the confinement potential, and consequently, the quantum dot
size. In Figs.~\ref{fig2}--\ref{fig4} these critical magnetic field values
are shown in a double-logarithmic plot as a function of the dimensionless
coupling constant $\lambda$. We display the magnetic field range between
$\ln \gamma = 0$ (that is, $\omega_c = \omega_0$) and $\ln \gamma = 3$
($\omega_c \approx 20\,\omega_0$).


Notice that plotted in a a logarithmic scale these functions are nearly
linear, besides the linearity is more pronounced for large momentum $M$
and large magnetic field values. This linearity of the above curves
\begin{equation}\label{power_a}
  \ln \gamma_M = a_M  + b_M \ln \lambda
\end{equation}
implies the simple power law dependence
\begin{equation}\label{power}
  \gamma_M \propto \lambda^{b_M}.
\end{equation}
We expect that such a plain dependence is a consequence of the
classical nature of the electron system in the quantum dot in the
limit of a strong perpendicular magnetic field (see e.~g.\ Ref.\
\onlinecite{dresden}) as we will show in the next section where a
simple quasi-classical description is developed.

We observe that the phase diagram corresponding to the two-electron
quantum dot (Fig.~\ref{fig2}) is more regular that the two others,
namely, the nearly-straight-line boundaries appear to be parallel and
the widths of the consecutive regions are decreasing monotonically.
In contrast, the plots pertaining to the cases of three
(Fig.~\ref{fig3}) and four (Fig.~\ref{fig4}) electrons include some
regions of considerably larger width which indicate an enhanced
stability of the corresponding ground states. In particular, such
regions correspond to the maximum density droplet states of
angular momentum $M = 3$ for three electrons and $M = 6$ for four
electrons.

Moreover, we see that at the lower right corner of Figs.\
\ref{fig3} and \ref{fig4} (three and four electrons, respectively)
corresponding to low magnetic fields and strong Coulomb coupling
there is a slight deviation from collinearity between neighboring
lines. In the case of the four-electron dot the phase boundaries
$\gamma_2$ and $\gamma_3$ even merge together indicating that for
$\ln \lambda > -0.3$ there is a direct transition from the state
with $M = 2$ into the state with $M = 4$. Similar transitions
where the angular momentum changes by $\Delta M = 2$ were also
found for higher values of angular momentum, namely, there are
transitions $10 \to 12$ and $12 \to 14$. These properties of the
four-electron quantum dot were already pointed out and discussed
in Ref.\ \onlinecite{maarten03}. As we will see shortly, the
quasi-classical model will overlook the presence of such
irregularities.

We note that in each of Figs.\ \ref{fig3} and \ref{fig4} an extra
phase boundary is present at low magnetic fields which was omitted.
These disregarded boundaries --- see Fig.\ 2(a) of Ref.\ \onlinecite{mikh3b}
and Fig.\ 3 of Ref.\ \onlinecite{maarten03} for phase diagrams of three-
and four-electron systems, respectively --- display a different trend.
They are of a purely quantum-mechanical nature and result from a specific
distribution of particles among the low-energy levels of a quantum dot,
i.~e.\ they are effects that can not be reproduced in a classical model.

\section{Quasi-classical description}

The 2D $N$-electron system confined by a parabolic quantum dot and
placed in a perpendicular magnetic field $B$ is described by the
following dimensionless Schr\"{o}dinger equation:
\begin{equation}\label{schred}
  \{H-E\}\Psi = 0,
\end{equation}
where
\begin{eqnarray}\label{ham}
  H &=& \frac{1}{2}\sum_{n=1}^N \left\{ \left(
  -i \nabla_n + \frac{\gamma}{2}[\mathbf{e}_z\times\mathbf{r}_n]\right)^2
  + r_n^2 \right\} \nonumber \\
  &&+ \sum_{n,m=1\atop n<m}^{N}\frac{\lambda}{|\mathbf{r}_n-\mathbf{r}_m|}.
\end{eqnarray}

We consider the ground state energy of the state of fixed angular momentum
$M$ as a function of Coulomb coupling constant $\lambda$ and the magnetic
field $\gamma$
\begin{equation}\label{energy}
  E = E(\lambda, \gamma, M).
\end{equation}
Solving the equation
\begin{equation}\label{enequ}
  E(\lambda, \gamma, M) = E(\lambda, \gamma, M+1)
\end{equation}
for $\gamma$ will define the above-introduced critical magnetic field
values $\gamma_M$ corresponding to the angular momentum transitions
from $M$ to $M+1$.

It is well known that in the strong magnetic field regime\cite{maks96}
the electrons form a Wigner crystal which for a system up to five
electrons is just a ring of equidistantly placed electrons.\cite{bed94}

Due to the rotational symmetry around the $z$-axis the Hamiltonian
(\ref{ham}) commutes with the total angular momentum operator
\begin{equation}\label{angmomoper}
  \hat{M} = -i\sum_{n=1}^N\frac{\partial}{\partial \varphi_n}.
\end{equation}
Therefore, the simplest way to calculate the ground energy of the
system is to exclude the magnetic field from the Schr\"{o}dinger
equation (\ref{schred}) by means of the transformation
\begin{equation}\label{wftrans}
  \Psi \to \exp\left(i\sum_{n=1}^{N}\varphi_n/N\right)\Psi,
\end{equation}
with the ensuing scaling of coordinates
\begin{equation}\label{scale}
  \mathbf{r}_n \to \mathbf{r}_n\left(1+\gamma^2/4\right)^{-1/4},
\end{equation}
which enables to transform the initial problem into the equivalent
problem
\begin{equation}\label{sredtrans}
  \{\tilde{H}-E_0(\lambda_0,M)\}\Psi = 0
\end{equation}
with the Hamiltonian
\begin{equation}\label{hamtrans}
  \tilde{H} = \frac{1}{2}\sum_{n=1}^N \left\{
  -\nabla_n^2 + r_n^2 \right\}
  + \sum_{n,m=1\atop n<m}^{N}\frac{\lambda_0}{|\mathbf{r}_n-\mathbf{r}_m|}.
\end{equation}
without a magnetic field.

The coupling constants and the eigenvalues of both problems
are related as
\begin{subequations}\label{rel}
\begin{eqnarray}
\label{rel1}
  \lambda &=& \lambda_0\left\{1 + (\gamma/2)^2\right\}^{1/4}, \\
\label{rel2}
  E(\lambda,\gamma,M) &=& E_0(\lambda_0,M)\sqrt{1+ (\gamma/2)^2}
  -\gamma M/2.\nonumber\\
\end{eqnarray}
\end{subequations}
Here, we again used our earlier convention whence the symbol $M$ stands
for the absolute value of the angular momentum. Note that the eigenvalue
$E_0(\lambda_0, M)$ indeed depends only on the absolute value of the
angular momentum, and Eq.\ (\ref{rel2}) is written in accordance with
our previous convention for $M$. Combining Eqs.\ (\ref{rel}) and
(\ref{enequ}) we arrive at the following equation
\begin{equation}\label{main}
  \frac{\gamma_M}{2\sqrt{1+ (\gamma_M/2)^2}}
  = E_0\left(\lambda_0, M+1 \right) - E_0\left(\lambda_0, M \right)
\end{equation}
which expresses the critical magnetic field values in terms of the
reduced problem (\ref{sredtrans}).

\section{Derivation of the power law dependence}

Now we turn to the solution of the eigenvalue problem defined by
Eq.\ (\ref{sredtrans}) which involves two peculiar features. One
is that we need to solve for a highly excited state with a large
angular momentum $M$ that corresponds to the ground state of the
original problem (\ref{schred}), and this fact complicates the
solution. Also, according to Eq.~(\ref{rel1}), in the strong
magnetic field limit Eq.~(\ref{sredtrans}) has to be solved for
small values of the coupling constant $\lambda_0$. Therefore, the
solution must be attainable by means of some perturbative technique.

Neglecting the electron-electron interaction term in the
Hamiltonian (\ref{hamtrans}) we obtain the zero order equation.
Due to the rotational symmetry the following radial equation
for the zero-order one-electron function can be written:
\begin{equation}\label{swf}
  \left\{-\frac{1}{2r}\frac{d}{dr}r\frac{d}{dr} + \frac{m^2}{2r^2}
  + \frac{1}{2}r^2 - \varepsilon_0(m)\right\}\psi_0(m|r) = 0.
\end{equation}
Here, the symbol $m$ stands for the single-electron angular momentum.
As this angular momentum is large the eigenvalue can be estimated by
approximating the effective potential
\begin{equation}\label{veff}
  V_{\mathrm{eff}}(r) = \frac{m^2}{2r^2} + \frac{1}{2}r^2
\end{equation}
by a parabolic potential in the vicinity of its minimum at
\begin{equation}\label{minimum}
  r_0 = |m|^{1/2}.
\end{equation}
Here we disregard the first-derivative term in Eq.\ (\ref{swf})
as its inclusion gives only negligible corrections in the
$m \to \infty$ limit.
This leads to the following estimate for the one-electron energy
\begin{equation}\label{oneground}
  \varepsilon_0(m) \approx V_{\mathrm{eff}}(r_0) = |m|,
\end{equation}
and the total zero-order energy of all electrons in the dot becomes
\begin{equation}\label{enzero}
  E_0 = N|m| = M.
\end{equation}

Let us now estimate the characteristic energies of the two possible
types of motion of the electrons: the radial vibrations and the motion of
electrons along the ring.

The energy of the radial vibrations can be obtained by expanding the
effective potential at the equilibrium point $r_0$. The second
derivative of the potential
\begin{equation}\label{vderiv2}
  V_{\mathrm{eff}}''(r_0) = 4
\end{equation}
is of order unity and does not depend on the orbital momentum
$m$. This fact enables us to disregard these vibrations as their energy
is small in comparison to the zero order energy $\varepsilon_0(m)$.
Moreover, these vibrations are the same for any orbital momentum $M$
(they are not affected by the electron-electron interaction which is
even smaller, i.~e.\  $\lambda/r_0 \ll 1$), and thus they can be
neglected when solving Eq.~(\ref{main}).

The energy of the longitudinal electron excitations along the ring is
even smaller. Actually, in zero order approximation the electron state
under consideration is degenerate with respect to these excitations
since the same total energy of the electronic ring (\ref{enzero}) can
be constructed from various angular momentum distributions among
individual electrons. Thus, this angular motion is highly affected
by the weak electron-electron interaction. The result of this effect
is well-know: it leads to the Wigner crystallization of the electrons
along the ring.

The first order correction to the eigenvalue
$E_0\left(\lambda_0, M\right)$ is obtained by including the
electron-electron
interaction term, but calculated classically for the electrons positioned
equidistantly on a ring of radius $r_0$. This energy correction reads
\begin{equation}\label{encor}
  \Delta E(\lambda_0, M) = \frac{\lambda_0}{r_0}f_N
  = \lambda_0\sqrt{\frac{N}{M}}\,f_N,
\end{equation}
where $f_N$ is the Coulomb energy of the equidistantly located
electrons on the unit radius ring, in particular,
\begin{equation}\label{coulombfact}
  f_2 = 1/2, \quad f_3 = \sqrt{3}, \quad f_4 = 1 + 2\sqrt{2}.
\end{equation}

Then, combining Eqs.~(\ref{enzero}) and (\ref{encor}) we obtain
the final expression for the energy to first order in powers
of $\lambda_0$
\begin{equation}\label{energyfin}
  E(\lambda_0, M) = M + \lambda_0\sqrt{\frac{N}{M}}\,f_N.
\end{equation}
Inserting this expression into Eq.~(\ref{main}), solving it to the
lowest accuracy in $\gamma^{-1}$ powers, and taking the average of
the difference of two inverse square roots we obtain the result
\begin{equation}\label{depfin_a}
  \gamma = \frac{2}{(Nf_N^2)^{1/3}}
  \left(M + \frac{1}{2}\right)\lambda^{-2/3}.
\end{equation}
Finally, taking the logarithm of Eq.\ (\ref{depfin_a}) and comparing
to Eq.\ (\ref{power_a}) we extract the expressions for the coefficients
\begin{eqnarray}\label{depfin_b}
  a_M &=& \ln (2M+1) - \frac{1}{3} \ln N - \frac{2}{3} \ln f_N, \nonumber\\
  b_M &=& -\frac{2}{3}.
\end{eqnarray}

\section{Comparison with the exact numerical results}

\begin{figure}
\includegraphics[width=80mm]{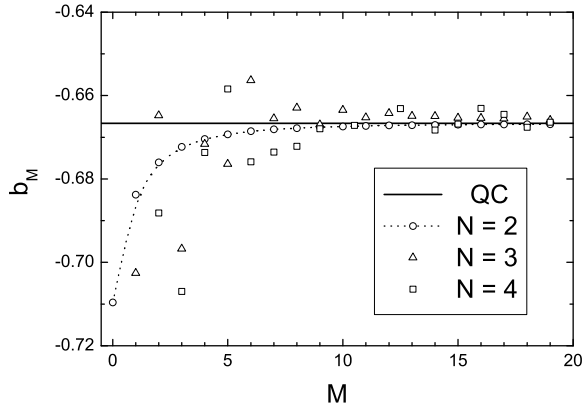}
\caption{The power law coefficients $b_M$ as a function of the
angular momentum $M$. The quasi-classical (QC) treatment predicts
a universal value $b_M = -2/3$ for any number of electrons shown
by the solid line. Different symbols indicate the exact values of
the coefficients for quantum dots containing $2$ to $4$ electrons.
The two-electron result displays a monotonic behavior and is
spline-approximated by the dotted line to guide the eye.}
\label{fig5}
\end{figure}

The quasi-classical solution predicts a universal value for the
power-law index $b_M = -2/3$ independent of the number of electrons.
The comparison with the exact diagonalization results is displayed
in Fig.~\ref{fig5}. As discussed above, in the four-electron dot
we encountered two cases when a single $\Delta M = 2$ transition
occurs in place of expected two $\Delta M = 1$ transitions in the
whole considered range of $\lambda$ and $\gamma$ values. In these
cases (namely, the transitions $10 \to 12$ and $12 \to 14$)
we used the $M$-value corresponding to the average of the two
expected $\Delta M = 1$ transitions.

In general, the discrepancy with our analytical model is small
as all results fall into a range between $-0.64$ and $-0.72$.
As expected, the deviation of the exact result from the
quasi-classical limit rapidly decreases with increasing angular
momentum $M$, and already at $M = 8$ the difference is less
than $2\%$.

The exact results corresponding to the two-electron dot show a
monotonous behavior as indicated by the dotted-line in
Fig.~\ref{fig5}. The reason is that the two-electron system
in a parabolic dot possesses only
one non-trivial degree of freedom as the centre-of-mass motion
can be separated from the relative motion and is not excited in
the ground state. In contrast, the internal motion of three-
and four-electron dots involves more degrees of freedom and
is more intricate. This results in a more complicated
dependence of $b_M$ on the angular momentum $M$. In particular,
enhanced non-monotonicities are discernible in the vicinity
of more stable maximum-density-droplet or ``magic'' states
at $M = 3$ and $6$ for three electrons and $M=6$ for
four electrons.

\begin{figure}
\includegraphics[width=80mm]{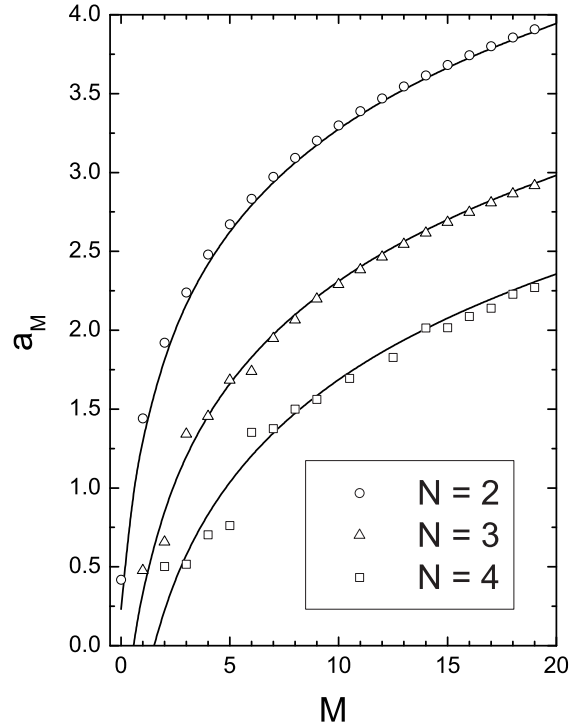}
\caption{The coefficient $a_M$ as a function of the angular momentum for
the case of $N = 2$, $3$ and $4$ electrons in the dot. The full lines
denote the quasi-classical (QC) result, and the symbols correspond to
the results from exact-diagonalization.}
\label{fig6}
\end{figure}

Fig.~\ref{fig6} shows the comparison for the coefficient
$a_M$. The quasi-classical model (\ref{depfin_b}) predicts
a logarithmic dependence of $a_M$ on the angular momentum $M$.
We see that for the two-electron system the correspondence
is nearly perfect.

The behavior concerning three and four electrons is more complex.
At low values of the angular momentum (and thus low magnetic fields)
there are notable discrepancies between the these sets of results
in Fig.~\ref{fig6}. In particular, we observe wide ``gaps'', i.~e.\
abrupt jumps in the $a_M$ values obtained from the exact-diagonalization
at angular momenta $M = 3$ ($M = 6$) corresponding to maximum density
droplet states in three (four) electron systems. Nevertheless, the overall
trend of these dependences still follows the quasi-classical prediction
rather closely, and the discrepancy between the classical and fully
quantum mechanical results vanishes very quickly with increasing $M$.

\section{Conclusions}

In conclusion, we presented a quasi-classical theory of the magnetic
properties of few-electron quantum dots whose main result is a
simple power law dependence of the critical magnetic fields on
the Coulomb coupling constant. While the main virtue of this theory
lies in its relative simplicity and ability to provide results
without resorting to heavy computation, the comparison of its
predictions to the exact results also proves its robustness even
in the realm of quantum mechanics.

\acknowledgments

This work is supported by the European Commission GROWTH program
NANOMAT project under contract No.\ GSRD-CT-2001-00545, the
Belgian Interuniversity Attraction Poles (IUAP), the Flemish
Science Foundation (FWO-Vl) and the Flemish Concerted Action
(GOA) programmes. E.A.\ is supported by the EU under contract
number HPMF-CT-2001-01195.

\end{document}